\documentclass[aps,amsmath,amssymb,
superscriptaddress]{revtex4}
 \usepackage{amsfonts}
\usepackage{graphicx}
\renewcommand{\>}{\rangle}
\newcommand{\<}{\langle}
\newcommand{\ba}{\begin{align}}
\newcommand{\ena}{\end{align}}

\begin{document}
\title{Dissipationless BCS Dynamics with 
Large Branch Imbalance}
\author{A. Nahum}
\affiliation{James Frank Institute, University of 
Chicago, 5640 S.
Ellis Ave. Chicago IL 60637}
\author{E. Bettelheim}
\affiliation{Racah Institute of Physics, The Hebrew 
University of
Jerusalem, Safra Campus, Givat Ram, Jerusalem, 
Israel 91904}
\begin{abstract}
In many situations a BCS-type superconductor will 
develop an imbalance 
between the populations of the hole-like and electron-like 
spectral branches. This imbalance suppresses the gap. 
It has been noted by Gal'perin, Kozub and Spivak \cite{spivak} 
that at large imbalance, when the gap is substantially 
suppressed, an instability develops. The analytic 
treatment of the system beyond the instability point 
is complicated by the fact that the Boltzmann approach 
breaks down. We study the short time behavior following 
the instability, in the collisionless regime, using methods 
developed by Yuzbashyan et al. \cite{kuznetsov, kuznetsov 2}.
\end{abstract}

\maketitle

\section{Introduction}

The excitation spectrum of a BCS superconductor 
consists of an electron-like and a hole-like branch. While 
the two
are equally populated in equilibrium, non-equilibrium
states may have a `branch imbalance'. For example, if a
superconductor is placed in an NSN junction as in the 
experiment of
Clarke \cite{clarke, tinkhamclarke, tinkham}, the injected 
quasi-particles can be primarily
electron-like, and the electron-like branch more heavily 
populated
than the hole-like branch in the steady state. 
Superconducting wires, 
where branch imbalance arises at phase-slip centres 
\cite{skocpol}, 
give another example.

This imbalance suppresses the spectral gap 
\cite{galaiko}. 
If large enough, it
returns the system to the normal state, which is 
known to be
unstable below $T_c$ -- this is the Cooper instability. 
It was
realized by Gal'perin, Kozub and Spivak \cite{spivak, 
spivak2} 
that the
Cooper instability is a limiting case of a more general 
instability
which afflicts any state beyond a critical level of
imbalance and results in oscillations of the order 
parameter. 
Traditional approaches to the
dynamics fail when the instability occurs, and for this 
reason the
supercritical behavior of the superconductor is as yet 
unknown. It
is important to resolve this, since situations in which 
large
imbalance occurs are quite natural, for example in NSN 
junctions
at large enough injection rate, or long superconducting 
wires
held at sufficiently high voltage.

This paper makes a step in this direction by treating the 
short-time 
dynamics
in the limit where dissipative processes act slowly in 
comparison
with the BCS dynamics (in particular the
oscillation of the gap). We work in the regime 
$T_c-T\ll T_c$, when
the slow relaxation of imbalance allows the system 
to reach a
`quasiequilibrium' whose deviation from 
equilibrium can be
characterized only by the amount of imbalance. We also 
assume the
initial conditions are close to the unstable stationary solution, 
in a sense 
defined below.

Since the BCS dynamics are characterized by a time 
scale of 
order $1/(T_c-T)$, the assumed separation of scales 
is 
$1/(T_c-T)\ll\tau_\epsilon$, 
where $\tau_\epsilon$ is the time scale associated 
with energy 
relaxation. In a metallic superconductor with Debye 
energy 
$\Theta\gg T_c$ and $\tau_\epsilon\sim\Theta^2/T_c^3$, 
this 
gives only the unrestrictive $(T_c-T)/T_c\gg T_c^2/\Theta^2$.

Often one can avoid the complexities of the microscopic BCS 
dynamics of a superconductor with a simpler effective 
description such as 
the Ginzburg-Landau or the Boltzmann kinetic equation. 
Two time 
scales are important in deciding whether either is
appropriate: the inelastic quasiparticle relaxation time,
$\tau_\epsilon$, and the time scale $\tau_\Delta$ over which 
the
order parameter varies significantly.

When $\tau_\epsilon\ll\tau_\Delta$, the quasiparticle 
distribution
rapidly reaches a local equilibrium characterized by the order
parameter $\Delta(\vec{r},t) = |\Delta(\vec{r},t)| e^{i
\chi(\vec{r},t)}$, with dynamics described by the 
Ginzburg-Landau
equations for $\Delta(\vec{r},t)$. We are interested in the 
opposite
limit, $\tau_\epsilon\gg\tau_\Delta$, which is usually tackled with
the Boltzmann kinetic equation \cite{aronov} for the quasiparticle
distribution function $n(\vec{r},\vec{p},t)$:
\begin{align} \label{Boltzmann}
& \frac{\partial n}{\partial t} + \frac{\partial \epsilon}{\partial
 \vec{p}}
\frac{\partial n}{\partial \vec{r}} - \frac{ \partial
\epsilon}{\partial \vec{r}} \frac{\partial n}{\partial     \vec{p}}=
-I\{n\}.
\end{align}
Here $\epsilon$ is the energy of a quasiparticle state and the 
functional $I\{n\}$ accounts for impurity
scattering and collisions between electrons or between 
electrons and phonons.
The kinetic equation must be supplemented with the self-
consistency 
equation, which determines
$|\Delta (\vec{r},t)|$ and thus the quasiparticle energies:
\begin{align}
\label{selfconsist1} &  1 = \frac{\lambda}{2} \int
\frac{1-2n_{\vec{p}}}{\epsilon (\xi_{\vec{p}})} d \xi_{\vec{p}}.
\end{align}
Here $\xi_{\vec{p}}=p^2/2m-\mu$, $\mu$ is the chemical 
potential,
and $\lambda$ the BCS coupling constant. In addition there is a 
`neutrality
condition' involving the
phase $\chi$ of the order parameter. This condition arises from 
the
continuity equation and ensures the conservation of charge -- it 
is independent of (\ref{Boltzmann}) in the case of a superconductor.
 In the
spatially homogeneous case, if we define $\Phi\equiv\frac{\hbar}{2}
\frac{\partial \chi}{\partial t} + e \varphi$ (where $\varphi$ is
the electric potential) and $\tilde{\xi}_{\vec{p}} = \xi_{\vec{p}} +
\Phi$, it takes the form
\begin{equation} \label{Phidef}
\int    \frac{ n_{\vec{p}  \phantom{|} }   \tilde{\xi}_{\vec{p}}}{\epsilon 
(\xi_{\vec{p}})}
d\xi_{\vec{p}}
=\Phi.
\end{equation}
This integral quantifies the branch imbalance. The electron- and
hole-like branches are distinguished by the sign of $\tilde\xi$. In
equilibrium the two branches are identically populated since the
quasiparticle energy ($\epsilon^2={\tilde\xi}^2+|\Delta|^2$) is even
in this quantity, and $\Phi$ vanishes.

It is important that while branch imbalance is absent in
equilibrium, its relaxation rate $\tau_Q$ diverges as $T_c$ is 
approached \cite{tinkham, schmidschon}: 
$\tau_Q\sim\tau_\epsilon T_c/\Delta$. 
 (This is in the absence of oscillations of the gap, and assuming
  relaxation due to electron-phonon collisions 
\cite{tinkham, aronov}.) Thus when
$\Phi\neq 0$, the quasiparticle distribution reaches a
`quasiequilibrium' on a time of order $\tau_\epsilon$, characterized 
by distinct chemical potentials for the separately equilibrated 
hole-like and electron-like
branches \cite{galaiko, tinkham, schmidschon, artemenko}. 
This distribution is given below (\ref{npwithimbalance}).
Inserting it into the self-consistency equation (\ref{selfconsist}) 
reveals that imbalance suppresses the gap relative to $\Delta_0$,
its value when $\Phi=0$:
\begin{equation}\label{gapsuppression1}
|\Delta|^2 =\Delta_0^2 - 2 \Phi^2.
\end{equation}
At $\Phi=\Delta_0/\sqrt{2}$, the gap is completely suppressed, 
and
the system is returned to the unstable normal state. The 
instability 
appears earlier, at an intermediate value $\Phi_c$.

The Boltzmann description cannot handle the system after the
instability takes hold, as modes are excited in which
Cooper pairs posses non-zero relative phases (the
relative phases of the $s_-(\xi)\equiv s_x(\xi)-is_y(\xi)$, in the 
language of Anderson
pseudospins \cite{anderson}). Only the overall phase of the 
condensate is
retained in the Boltzmann approach, effectively restricting 
the system to a subclass of solutions where the Cooper pairs 
precess in phase.

In this situation we must return to the Gorkov equations 
describing
the mean-field dynamics of the individual Cooper pairs
\cite{volkovkogan}. We study here only the limit in which
dissipative processes are neglected
($\tau_\epsilon\rightarrow\infty$) and the dynamics controlled
purely by the BCS Hamiltonian. This was done for the Cooper 
instability by Barankov, Levitov and Spivak in \cite{bls}, 
yielding a `soliton train' of peaks in the gap. The problem 
was further discussed by Warner and Leggett in \cite{leggett},
 by Barankov and Levitov in \cite{bl, bl2}, and by Yuzbashyan 
 and Dzero in \cite{dynamicalvanishing}. The
integrability of the mean-field BCS system was established by
Yuzbashyan, Altshuler, Kuznetsov and Enolskii in the papers
\cite{kuznetsov, kuznetsov 2} and a general framework for 
addressing
its dynamics was developed, which we use here to tackle the 
more
involved case of imbalance. We confine ourselves to the 
integrable BCS Hamiltonian because it can be treated 
analytically; it was however shown by Barankov and Levitov
 in \cite{levitovhiggs} for the case of the Cooper instability 
 that the gap oscillations survive the breaking of integrability.

Our results show the emergence of oscillatory behavior at the
instability point. On much larger time scales ($\sim 
\tau_\epsilon$)
dissipation will modulate the form of these oscillations and
determine the final fate of the system. Such an analysis is 
beyond
the scope of this paper, but the solutions we present may 
be thought of as candidates to be found either at very
large times after the onset of the instability or at intermediate 
times
on the way to the asymptotic behavior, and are a first step in
 calculating the long-time behavior. We discuss this
  briefly in Section \ref{conclusion}.

The ability to `switch on' the pairing interaction in ultracold
trapped gases via tunable Feshbach resonances 
\cite{f1,f2,f3,f4,f5,f6} means that one might hope to 
observe the oscillations of the order parameter 
directly in the case of the Cooper instability  
\cite{bls,leggett,dynamicalvanishing,bl2,bl}. 
On the other hand, it is not clear whether one could create
 imbalance in a sufficiently controlled fashion to observe
  the instability it creates\footnote{
Population imbalance, i.e.  having more spin-up than 
spin-down electrons, may be more easily created. The 
dynamics for this distinct situation is addressed in 
\cite{populationimbalance}.
}. In contrast, imbalance can be created easily in metals, 
for example in tunnel junctions \cite{clarke, tinkham,
tinkhamclarke}. Here the direct observation of the 
collisionless dynamics is unlikely due to short 
dissipation times, and its significance is instead in 
its effect on processes at longer timescales such 
as the relaxation of imbalance.

The structure of the paper is as follows. Section 
\ref{gorkoveqns}
presents the equations of motion for the system 
and the relevant
initial conditions, and includes the final result of the 
analysis.
Section \ref{elements} describes the formal 
solution to BCS dynamics
\cite{kuznetsov, kuznetsov 2, relaxationandpersistent, 
erratum},
which is applied to the relevant case in Section
\ref{solutionwhereimbalance}. We conclude in Section
\ref{conclusion}.

\section{Gorkov's nonlinear equations \& solution with 
imbalance}
\label{gorkoveqns}
The system of equations describing the BCS superconductor 
in the
non-dissipative regime were derived by Volkov and Kogan
 \cite{volkovkogan} in the Keldysh Green's function formalism:
\begin{eqnarray}\label{VolkovKogan}
\dot{\vec{s}}(\xi) = {\vec{s}}(\xi)\times  (2 \Delta_x , 2 \Delta_y
, - 2 \xi),
\end{eqnarray}
where $\vec{s}$ is defined by the following Keldysh Green's
functions:
\begin{eqnarray}
\label{keldysh}
s_z (\xi) =    \< [c_\uparrow(\xi),  c^\dagger_\uparrow(\xi) ] \>
\nonumber\\
s_- (\xi) =    \< [c_\uparrow(\xi), c_\downarrow(\xi)  ] \>
\end{eqnarray}
$s_- = s_x - i s_y$, and:
\begin{eqnarray}\label{selfconsist}
\Delta = \Delta_x - i \Delta_y = \frac{\lambda}{2} \int s_-(\xi) 
d\xi.
\end{eqnarray}
This system of equations can be derived from a classical 
Hamiltonian:
\begin{eqnarray}
 H = \int 2 \xi s_z(\xi) d\xi - \frac{2}{\lambda}|\Delta|^2; \qquad 
 \{s_i(\xi), s_j(\xi')\}= \epsilon_{ijk} s_k(\xi)\delta(\xi-\xi').
\end{eqnarray}
This is the mean-field BCS
Hamiltonian written in terms of Anderson pseudo-spins, with 
an up (resp. down) spin representing a full (empty) Cooper pair. 
Singly-occupied pairs decouple from the order parameter 
dynamics (as can be seen from the BCS Hamiltonian which 
involves only pair-to-pair scattering) and correspond to zero-length
 spins. One may derive the Gorkov equations heuristically as
a mean-field approximation to the BCS Hamiltonian.

The current paper aims to present solutions of (\ref{VolkovKogan}) 
for imbalanced superconductors,
using the approach of \cite{kuznetsov}. Before
doing so, let us recall the instability of the stationary
solutions of (\ref{VolkovKogan}) in the presence of imbalance which
 was pointed out in \cite{spivak}.
Generally, stationary solutions of (\ref{VolkovKogan}) have the
form:
\begin{eqnarray}\label{stationaryfg}
s_z(\xi) &=& -\frac{\xi + \Phi}{\sqrt{(\xi+\Phi)^2 + |\Delta|^2}} 
(1-2 n(\xi))\\
s_-(\xi)  &=& \phantom- \frac{\Delta}{\sqrt{(\xi+\Phi)^2 + |\Delta|^2}} 
(1 - 2
n(\xi))\nonumber
\end{eqnarray}
where $n(\xi)$ is the quasi-particle distribution function, which in
the presence of imbalance is\footnote{
These initial conditions (which will later be perturbed) are thermal 
averages. Typical initial conditions, as opposed to averaged initial 
conditions, need not have an $n(\xi)$ that is smooth on the scale 
of the level spacing. Justification for the use of thermally averaged 
initial conditions is given in \cite{bl2}: we can average our spins 
over a small energy range containing many levels to give a smooth 
$n(\xi)$ without changing the equations of motion 
(\ref{VolkovKogan}, \ref{selfconsist}).
} \cite{tinkham, spivak}
\begin{eqnarray}\label{npwithimbalance}
n(\xi) = \frac{1}{\exp\left(\frac{\sqrt{(\xi+\Phi)^2 + |\Delta|^2} -
\Phi\, \mbox{sign}(\xi + \Phi)}{T}\right) + 1}
\end{eqnarray}
which is a Fermi-Dirac distribution for the quasi-particles, with 
different chemical potentials for the hole-like and electron-like
 spectral branches implemented by the term $\Phi\,\mbox{sign}
 (\xi  +
\Phi)$ in the exponent. $\Phi$ parametrizes the amount of 
imbalance in the system. The expression for $n(\xi)$ is valid
 when $|\xi+\Phi| \gg |\Delta|$. The self consistency condition
  (\ref{selfconsist}) is satisfied if \cite{galaiko}:
\begin{equation}\label{gapsuppression}
|\Delta|^2 =\Delta_0^2 - 2 \Phi^2
\end{equation}
where $\Delta_0$ is the order parameter at the temperature 
$T$ in the absence of imbalance (for the case $\Phi=0$). 
Thus imbalance between electron-like and hole-like excitations 
suppresses the order parameter\footnote{To obtain 
(\ref{gapsuppression}) one solves the self-consistency 
equations with (\ref{npwithimbalance}). The self-consistency 
condition is dominated by spins in the energy range $|\xi+\Phi| 
\sim T_C$, where (\ref{npwithimbalance}) is valid.}.

One can easily see that the distribution is unstable when $\Phi
 = \Delta_0/\sqrt{2}$, and $\Delta=0$ by
(\ref{gapsuppression}). At this point it becomes
\begin{eqnarray}\label{normalindisguise}
n(\xi) = \frac{1}{\exp \left(\frac{\xi \,
\mbox{sign}(\xi+\Phi)}{T}\right) +1}
\end{eqnarray}
which is nothing but the quasi-particle distribution of a normal 
Fermi gas in
the excitation representation, where an artifical distinction 
between hole-like and electron-like excitations is made at 
$\xi=-\Phi$.
So, the peculiar form of (\ref{normalindisguise}) is an artifact
of the excitation representation, and it just describes a normal
 metal placed
at $T<T_c$. The Cooper instability of this metal
presents itself as a instability of the stationary solution of
equation (\ref{VolkovKogan}).

At $\Phi=0$ however the solution is stable, and represents 
the equilibrium superconducting
state. There must therefore be an onset of instability at some 
finite $\Phi$
intermediate between $0$ and $\Delta_0/\sqrt{2}$. In order to 
find
this point and in order to give a quantitative
characterization of the instability, a linear stability analysis was
performed around this solution in \cite{spivak}, looking for the 
presence of an unstable mode $e^{-i\omega t}$, $\mbox{Im}
(\omega)>0$. This leads to an integral equation for $\omega$:
\begin{eqnarray}
G(\sqrt{\omega^2/4-|\Delta|^2}) \equiv \int \frac{d\xi}{\sqrt{
\tilde\xi^2+|\Delta|^2}} \frac{1-2 n(\xi)}{\tilde\xi-
\sqrt{\omega^2/4-|\Delta|^2}} =0.
\end{eqnarray}
Here $\tilde\xi=\xi+\Phi$. The solutions of this equation were 
found to be:
\begin{equation} \label{Gu=0}
\mbox{Re} (\omega) = 2 \Phi,\qquad |\mbox{Im} (\omega)|=
 \frac{2}{\pi}((T - T_c) -
|\Delta|).
\end{equation}
The instability arises when the right hand side of this expression
 for $|\mbox{Im} (\omega)|$ becomes positive at $\Phi_c$. 
 When the imbalance completely suppresses the gap 
 (\ref{Gu=0}) agrees well with the usual Cooper instability case.
  In fact all we shall need in the following are the orders of 
  magnitude of the the following quantities:
\begin{align} \label{orderofmag}
\Delta\sim s^2 T_c ,\qquad \gamma\equiv\frac{1}{2} \mbox{Im}
 (\omega) \sim s^2 T_c, \qquad \Phi \sim s T_c
\end{align}
where $s\equiv \sqrt{(T_c-T)/T_c}$, and we have also defined 
the parameter $\gamma$, the rate of instability, an important 
parameter which will appear frequently below. The order of 
magnitude of the different quantities could just as well have 
been taken from the Cooper instability case, as they remain 
the same when the imbalance completely suppresses the 
gap. Namely these orders of magnitude are not sensitive to 
the exact form of the distribution function, but rather to gap 
suppression  and may be viewed as consequences of 
Eq. (\ref{gapsuppression}).

\begin{figure*}
\centering{
\includegraphics[width=3.0in]{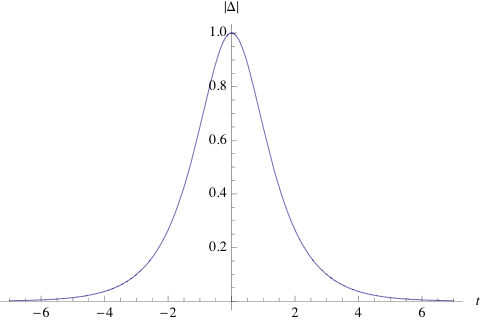}
\includegraphics[width=3.0in]{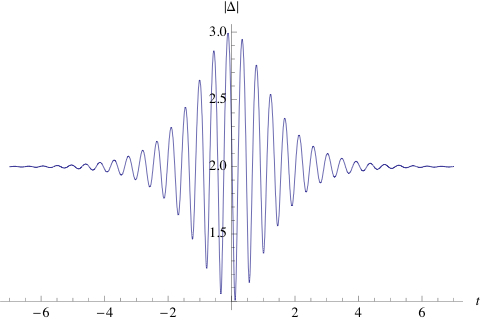}
}
 \caption{The soliton for the Cooper instability of a normal metal, 
 with $\gamma=0.5$, and a pulse of oscillations of $|\Delta(t)|$ for 
 $\Phi=7$, $\gamma=0.5$, $\Delta=2$, showing the qualitative 
 nature of the solution in the two regimes $\Delta=0$ and 
 $2\gamma<\Delta$. The solution interpolates smoothly 
 between these cases.}
  \label{overView}
  \end{figure*}

The main result of the following analysis is the time dependence
 of the order parameter following a perturbation from the 
 unstable stationary solution. This is a train of `solitons' 
 each of the form:
\begin{align}
\label{gap}
\Delta(t)= \Delta
\left(1+
\frac{2\gamma}{\Delta}e^{-2 i \Phi (t-\chi)}\text{sech} 
{(2\gamma t)}
\right)
\end{align}
where $\Delta$ (as opposed to $\Delta (t)$) denotes the 
value of the order parameter, which we can take to be 
real and positive, in the stationary but unstable initial state. 
The solitons are separated in time by a period
that increases logarithmically with the smallness of 
the perturbation,
\begin{equation}
t_0=\gamma^{-1}\log\left(\frac{4\gamma}{\delta}\right)
\end{equation}
in the notation used below. $\chi$ is a real constant 
(see below) which does not affect the qualitative nature 
of the soliton.

The behaviour of $|\Delta(t)|$ consists in fast (at frequency 
$2\Phi$) ocillations within an envelope given by 
${|1\pm(2\gamma/\Delta)\text{sech} {(2\gamma t)}|}$. 
When $2\gamma/\Delta\gg 1$ (recall that as 
$\Delta\rightarrow 0$ we return to the finite-temperature 
normal metal) this formula gives a soliton of the same 
shape as that found in \cite{bls} for the order parameter 
oscillations associated with the Cooper instability. In 
the opposite limit where the instability is small and 
$2\gamma/\Delta\ll 1$, we find instead pulses of 
oscillations of $|\Delta(t)|$ during which the value of 
$|\Delta(t)|$ averaged over over a period of the 
fast oscillation rises: $|\Delta(t)|_{av}\simeq\Delta 
(1+\gamma^2/\Delta^2 \text{sech}^2 (2\gamma t))$.

\section{Elements of the integrable structure}
\label{elements}

The instability discovered by Gal'perin et al. shows 
that at
large times dissipation may take the system to a new
state which is very different to the initial stationary 
solution. A similar situation exists when a normal metal 
is
placed at $T<T_c$. For comparison we briefly review 
here the known
results on this instability, the current paper expounding 
on these
results to include the case where imbalance is present.

A normal metal at $T<T_c$ is unstable against the 
development of a gap (this is the celebrated Cooper 
instability). The short term dynamics of the
system, just after the instability takes hold, consist in an
oscillatory behavior of $\Delta$. These oscillations damp 
out at
large times because of dissipation (this process is not 
described by the pure BCS Hamiltonian) and the material 
is left in the superconducting state.
The oscillatory behavior of
$\Delta$ consists of a `soliton train' which can
be found by solving (\ref{VolkovKogan}) for a system 
placed near the
normal metal state -- namely with initial conditions
(\ref{stationaryfg}), where $n_p$ is the Fermi distribution 
function
plus a small perturbation. This is discussed in \cite{bls,
leggett,bl,bl2}.

The soliton train describes the short-time 
($\ll \tau_\epsilon$) behavior of the order parameter 
following the onset of the Cooper instability. A similar 
analysis will
be presented here for the case when the initial state is 
given by (\ref{stationaryfg}) and (\ref{npwithimbalance}), 
i.e. a superconductor with branch imbalance near $T_c$. 
We study the behavior of the system after a small 
perturbation is
added to $n(\xi)$. This describes the short time behavior 
after the
instability has taken hold. The approach also yields 
solutions where
the perturbation is rather larger. We believe that these 
solutions
may provide further intuition about the possible routes 
the system
may take once dissipation effects are taken into account.

In order to find these solutions we apply the formalism 
developed in \cite{kuznetsov, kuznetsov 2} for the 
dynamics of the mean field BCS system, which draws 
heavily on the
theory of integrable systems. We shall establish the 
notation and
present the main concepts of the derivation, referring
 the reader interested in further details to the original
  papers. In these
papers the spectrum is treated as discrete -- here we 
assume a continuous spectrum.

An important object in the integrable structure is
the Lax matrix of the system, a $2\times 2$ matrix 
depending on a complex
parameter $u$, given by:
\begin{eqnarray}\label{laxmat}
\mathcal{L}(u)=\left(
\begin{array}{cc}
A(u) & B(u) \\
B^*(u) & -A(u)
\end{array}
\right)
\end{eqnarray}
where:
\begin{align}
& A(u) = \frac{2}{\lambda} - \int \frac{s_z(\xi)}{u-\xi} 
d\xi, \\
& B(u) = \int \frac{s_-(\xi) }{u-\xi} d\xi \nonumber.
\end{align}
For any $u$, the matrix $\mathcal{L}(u)$ is time 
dependent via $\vec s$, but its eigenvalues are 
not -- a key to integrability. These eigenvalues
constitute an infinite set of constants of 
motion\footnote{In the
case where the spectrum is discrete only a 
finite number of these
constants of motion are actually independent.}. 
They are
labeled $v(u)$, and given by:
\begin{align}
\label{specpol}
v(u) = \pm \sqrt{-\det\mathcal{L}(u)}.
\end{align}
The analytic structure of $v(u)$ is as follows: in
 general $v(u)$ has branch cuts, with square 
 root behavior around the branch
points, parallel to the imaginary $u$ axis, as
 well as a jump
discontinuity on the real axis, on the support of 
the spectrum. The
branch points $E_i$ satisfy $v(E_i)=0$. Other
 important quantities are the zeros of $B(u)$, 
 dubbed $u_i$:
$B(u_i)=0$.

An important simplification takes place if one
 is interested only in
the long time behavior of the system (but still 
at times $\ll \tau_\epsilon$). After an initial
 transient the
system exhibits periodic or quasi-periodic
 behavior whose frequency
is dictated by the branch points of $v(u)$. 
The jump
discontinuity is only relevant for the initial
 transient. The
oscillatory behavior following the transient 
is captured by a
simpler system containing only a finite number 
of spins
$\vec{s}^{(i)}$, $i=1,...,g+1$, where $g+1$ is 
the number of branch cuts
in $v(u)$. The system with a finite number of 
spins is integrable,
with a similar integrable structure to the infinite 
system,
integrals being replaced by sums; namely if
\begin{align}
\label{3spin}
& A(u) = \frac{2}{\lambda} - \sum_{i=1}^{g+1}
 \frac{s_z^{(i)}}{u-\xi_i}  \\
& B(u) = \sum_{i=1}^{g+1} \frac{s_-^{(i)} }
{u-\xi_i}\nonumber
\end{align}
are substituted into (\ref{laxmat}) then the 
eigenvalues of the
matrix are constants of motion. It is convenient
 to define $P(u) \equiv
\prod_i (u - \xi_i)$, and also a degree $2g+2$
 polynomial $Q(u)$ with zeros at
$E_i$, through $v(u) = \frac{2}{g}\sqrt{Q(u)}/P(u)$.
 The
appearance of $\sqrt{Q(u)}$ signals the relevance
 of the algebraic
Riemann surface defined by the curve $y(u) = 
\sqrt{Q(u)}$ to this
problem. To find the configuration of the finite 
number of spins at
any given time one must know $Q(u)$, which is 
independent of time,
and the time-dependent quantities $u_i(t)$,
 defined by
$B(u_i(t))=0$.

\begin{figure*}
\centering{
\includegraphics[width=95mm]{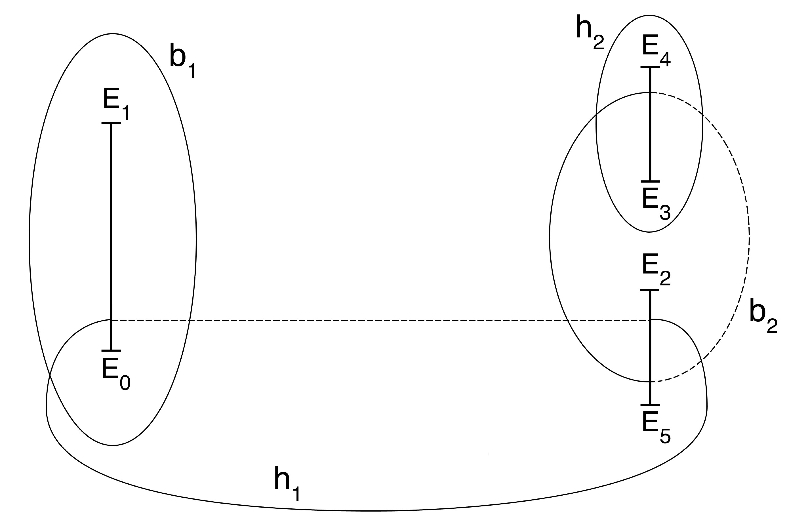}
}
 \caption{The three-sheeted Riemann surface with
branch points and cycles labelled.} \label{cyclesfig}
  \label{overView}
  \end{figure*}

To find the dependence of the $u_i$ on time, it 
is best to make
use of the connection of integrable systems and 
Riemann surfaces. A
central theme in the study of Riemann surfaces are the 
cycles, which are the
non-trivial closed curves on the Riemann surface (those that 
cannot
 be smoothly shrunk to a point). They will be denoted by $b_k$ 
 and
  $h_k$, $k=1,...,g$. These
cycles are depicted in Fig. \ref{cyclesfig}. Another mainstay of
the theory of Riemann surfaces are the Abelian (meromorphic) 
differentials
 on the surface. The so-called differentials of the first kind, which 
 are
everywhere holomorphic, form a $g$ dimensional vector space 
for
 which a basis is:
\begin{align}
\hat{u}_k = \frac{d u}{\sqrt{Q(u)}} u^{k-1}\quad k=1,\dots,g.
\end{align}
The differentials may be integrated around the non-trivial cycles to
assist in the following definitions:
\begin{align}\label{cyclesdefinitions}
\omega_{k,l} =\frac{1}{2} \oint_{b_k} \hat{u}_l,\quad \omega'_{k,l} =
\frac{1}{2}
\oint_{h_k} \hat{u}_l,\quad \tau = \omega^{-1} \omega'.
\end{align}
A familiar construction for genus one Riemann
surfaces, which have the topology of a
torus, allows us to represent the surface as a
rectangle with opposite edges identified. The rectangle is
characterized by its aspect ratio, which is an invariant of the
Riemann surface as well. In the case of genus one the aspect 
ratio 
of
the rectangle turns out to be equal to $-i \tau$ from
(\ref{cyclesdefinitions}). When the genus is higher than one we
encounter a matrix, $\tau$, which is a
generalization of the number $\tau$ of the genus-1 case. The
rectangle with opposite sides identified also has an analogue for 
higher genera: it is replaced by $2g$ (real) dimensional volume in 
$\mathbb{C}^g$ given by $\mathbb{C}^g /
(\mathbb{Z}^g \omega + \mathbb{Z}^g \omega')$. This $2g$ 
dimensional
volume is an analog of the $2$ dimensional rectangle in the 
genus-1
case, in that there exists an invertible map taking sets of points on 
the Riemann
surface into it. This is given by $\vec{J}(\{u\}): \Omega
\to \mathbb{C}^g / (\mathbb{Z}^g \omega + 
\mathbb{Z}^g \omega')$,
where $\Omega$ is the space of sets of $g$ 
points $\{ u_1,\dots,u_g\}$
(these are not ordered sets, so permutations are 
considered equivalent):
\begin{align}
J_j\left(\{u_i\}_{i=1}^g\right) = \sum_{i=1}^g \int_{P_0}^{u_i}
d\hat{u}_j.
\end{align}
The space  $\mathbb{C}^g / (\mathbb{Z}^g \omega +
 \mathbb{Z}^g
\omega')$ is called the Jacobian. The contour of integration 
from the
arbitrary initial point $P_0$ to the point $u$ can wind around 
any
of the cycles any number of times, so as a mapping to 
$\mathbb{C}^g$ it is only defined up to the addition of an 
element of the lattice
$\mathbb{Z}^g \omega + \mathbb{Z}^g \omega'$. This is
 however enough to give a well-defined map to
  $\mathbb{C}^g / (\mathbb{Z}^g \omega +
\mathbb{Z}^g \omega')$. One can check that in the case 
of genus one
the mapping takes a point on the Riemann surfaces and
 maps it onto a
rectangle whose aspect ratio is $-i\tau$, the rectangle being
represented by $\mathbb{C} / (\mathbb{Z} \omega + 
\mathbb{Z}
\omega')$.

The concept of the Jacobian is particularly important in the
solution of the problem because one can show that the 
zeros of
$B(u)$, which are denoted by $u_i$, satisfy the equation:
\begin{align}
\vec{J}(\{u_i(t)\}_{i=1}^g) = (c_1,c_2,\dots,c_g+2it)
\end{align}
where we have now written the time dependence of $u_i$ 
explicitly, while the $c_i$ are defined by:
\begin{align}\label{c}
c_i =  \sum_{j=1}^g \int_{E_{2j}}^{u_j(t=0)} d\hat{u}_i.
\end{align}
The roots $E_i$ of $Q(u)$ are listed for our case in 
(\ref{Es}). Considered as constants of integration for the 
dynamics of the $g$ roots $u_i(t)$ of $B(u)$, the $c_i$ 
are $g$ free complex variables determining the initial 
values $u_i(0)$. However, not all initial values are 
permissible, i.e. an arbitrary $\vec c$ will not correspond 
to a configuration of the spins; there are $g$ constraints 
on $\vec c$.

The $u_i$ together with the spectral curve $Q(u)$ contain 
all the
information needed to find the configuration of the spins. 
The
problem of finding the $u_i$ is thus the problem of 
inverting the
map $\vec{J}$. This is a solved mathematical problem 
with a long history,
which goes by the name `the Jacobi Inversion problem' 
\cite{enolskii}. The
solution can be obtained in terms of the Riemann 
$\theta$-function. We are interested here in the order 
parameter, whose
logarithmic derivative in time is given by $\sum_i u_i$.
 The
explicit solution of the inversion problem for BCS 
dynamics is presented in \cite{kuznetsov}, and 
gives the
time dependence of the order parameter as:
\begin{align}\label{thetafunctionexpression}
\Delta(t) =
\frac{\lambda}{2}\sum_{i=1}^{g+1} s^{(i)}_-=
C \exp{\left( 2 \vec{d} \eta  (\omega^T)^{-1} \vec{x}-
i\beta t\right)}
\frac{\theta\left((2\omega^T)^{-1}\left(\vec{x}+\vec{d}
\right)\left|\tau\right.\right)}
{\theta\left((2\omega^T)^{-1}\left(\vec{x}-\vec{d}
\right)\left|\tau\right.\right)}
\end{align}
provided that $\eta$ and $d$ are given by:
\begin{align}
\label{d}
\eta_{k,l} = - \sum_{j=l+1}^{2g+2-l} \frac{j-l}{4 (j+l)!}
\frac{d^{j+l} Q(u)} {d u^{j+l}}  \omega_{k,j},\qquad d_j 
=
\int_{E_0}^\infty d\hat{u}_j.
\end{align}
The frequency $\beta$ can be written in terms of 
the roots $E_i$ of $Q(u)$ as $\beta=2 \sum_i E_i$.

\section{Solution where imbalance is present}
\label{solutionwhereimbalance}

In the stationary (but unstable) imbalanced state, 
$s_z(\xi)$ and $s_-(\xi)$ are given by
the expressions (\ref{stationaryfg}) and
 (\ref{npwithimbalance}). In
this case, the eigenvalue $v(u)$ is given by:
\begin{align}
v^2(u-\Phi) &= (u^2 + \Delta^2) G(u)^2.
\end{align}
Throughout we use $\Delta$ (as opposed to 
$\Delta(t)$) to denote the value of the gap before 
the perturbation. $G(u)$, which was defined in 
(\ref{Gu=0}) because it appeared in the linear 
stability analysis, has appeared again as a common
 factor in $\det \mathcal{L}(u)=-A^2(u)-B(u)B^*(u)$.
 This is not a coincidence -- the connection between 
 roots
  of $v(u)$ and modes present in the solution is 
  explained 
  further in \cite{relaxationandpersistent}. Since 
  $v(u)^2$ 
  has six roots we are led to consider the three spin 
  problem, which
according to the arguments of the previous section 
represents
the dynamics of the order parameter after an initial 
transient. The polynomial
$Q(u)$ for the three-spin problem has the same 
roots as $v(u)^2$, given thus by:
\begin{align}
\label{shiftedQ}
Q(u) = (u^2 + \Delta^2) (u-\Phi - i \gamma)^2
 (u-\Phi + i \gamma)^2.
\end{align}
(Here we have shifted $u$ by $\Phi$ -- from the 
equations of motion in \cite{kuznetsov} we see that 
this can be compensated by giving the order 
parameter an additional phase factor.)
$Q(u)$ has a very particular structure, related to the 
fact that it
is a stationary solution. It has single roots only at 
$\pm i \Delta$
and the rest of its roots are double roots. This 
insures that the
Riemann surface given by $Q(u) = y^2(u)$ is of genus $0$.

We now wish to perturb the initial conditions. Unless
we fine-tune the perturbation to avoid doing so, we
will lift the degeneracy associated with the double
roots: the polynomial
$Q(u)$ will now have $6$ single roots (which must still 
occur in complex conjugate pairs, since $Q$ has real coefficients):
\begin{align}
Q(u) = (u^2 + \Delta^2)\left( (u-\Phi - i \gamma)^2 + \delta^2 \right)
\left( (u -\Phi+ i \gamma)^2 + \delta^2 \right).
\end{align}
For the integrals in (\ref{c}), (\ref{d}) we need the definitions (Fig. 
\ref{cyclesfig}):
\begin{align}
\label{Es}
(E_0,E_1,E_2,E_3,E_4,E_5)=(-i\Delta,i\Delta,\Phi-i \gamma+i\delta,
\Phi+i\gamma-i\delta,
\Phi+i\gamma+i\delta,\Phi-i\gamma-i\delta).
\end{align}
We neglect here (for example) the small shift in the position of the 
pair of roots around the origin: whereas the splitting of the double 
roots has a qualitative effect, such shifts have negligible effect on 
the solution, involving only small shifts in the parameters $\Delta$, 
$\gamma$ and $\Phi$, and a small change in the overall rate at 
which the phase of the order parameter rotates. Also, while in 
general $\delta$ can be complex, in the regime of interest to us 
the solution is not sensitive to the phase of $\delta$ except 
through the value of $\vec c$. In the following we treat $\delta$ 
as real unless otherwise stated.

 After the perturbation the
Riemann surface is of genus two and $\Delta(t)$ is given by 
(\ref{thetafunctionexpression}) with genus two hyperelliptic 
$\theta$-functions. The expression is quite
formidable, yet certain features can be clarified without much
analysis. Most notably, the solution
is quasi-periodic, with quasi-periods which can be deduced
straightforwardly from the general periodicity properties of 
$\theta$
-functions together with the particular form the matrices 
$\omega$
and $\tau$ take in this case. 

Before continuing to the analysis of the small perturbation case, 
we
note that if the perturbation is large enough it can lead to the
appearance of new roots for $Q(u)$, and to higher spin solutions
(with higher genera) -- this case is too general for us to say much 
about.

We assume that the parameter $\delta$ is small, as discussed in 
Section \ref{conclusion}. We then take the leading order in 
$\delta$ of
the expressions for the matrices $\tau$ and $( 2\omega^T )^{-1} 
(\vec{x} \pm \vec{d} )$, which figure in
(\ref{thetafunctionexpression}). We also expand in 
$s=\sqrt{(T_c-T)/T_c}$, taking the lowest order terms for each 
element. Then we have, to leading order (in practice we must 
make sure lower order terms do not contribute):
\begin{align}
\label{omega}
\omega^{-1}= \left(\begin{array}{cc}
-\frac{i\Phi^2}{\pi} & \frac{i\gamma\Delta^2 }{\Phi} \log^{-1}\left(
\frac{4\gamma}{\delta} \right)  \\
 \frac{i\Phi}{\pi}&
i\gamma \log^{-1}\left(\frac{4\gamma}{\delta}\right)
\end{array}\right)
;\qquad
\eta= \left(\begin{array}{cc}
0&0  \\
-\frac{i\pi\Delta^2}{2\Phi}&
\frac{i\Phi^2}{2\gamma}\log\left(\frac{\delta}{4\gamma} \right)
\end{array}\right)\nonumber;
\end{align}

\begin{align}
\tau &= \left(\begin{array}{cc}
\frac{i}{2\pi} \log \left( \frac{4 \Phi^4}{\gamma \delta \Delta ^2} 
\right) & -\frac{1}{2}
+ \frac{i\gamma}{\Phi}\log^{-1}{\left( \frac{4\gamma}{\delta}
\right )} 
\\ -\frac{1}{2} + \frac{i\gamma}{\Phi}\log^{-1}{\left( \frac{4
\gamma}
{\delta}\right )}& \frac{i\pi}{2} \log^{-1} \left(\frac{4
\gamma}{\delta} \right)
\end{array}\right)\nonumber;
\\ \nonumber\\
(2\omega^T)^{-1} (\vec{x} \pm \vec{d}) &= \left(
\begin{array}{cc}
- \frac{  \Phi t}{\pi} \pm \frac{i}{2\pi} \log\left( \frac{i \Delta}{2 
\Phi}
\right)
,\left(
\gamma t  \pm
\frac{i \gamma}{2\Phi}
\right)  \log^{-1} \left( \frac{\delta}{4\gamma}\right)
\end{array}
\right)+(2\omega^T)^{-1} \vec{c}.
\end{align}
We will return to the vector $\vec{c}$, which requires further 
analysis. The period matrix $\tau$ is seen to have very large 
and very small
elements on the diagonal, diverging or vanishing with
$\log^{\pm1}(\delta)$. Because of this the $\theta$-function is 
well
approximated by trigonometric functions. First we use the 
modular invariance
of  $\theta$-functions to trade in our period matrix for one 
whose
elements are all of order
$\log(\delta)$. This is done via the identity:
\begin{align}
\theta \left(\vec{y}  \left|
\left(\begin{array}{cc} i A & -\kappa \\
-\kappa & i h \end{array}\right)
\right)\right.=
\frac{e^{-\frac{\pi}{h}y_2^2}}{\sqrt{h}}
\theta \left(
 \left(\begin{array}{cc} y_1-\frac{i \kappa}{h}y_2 \\ -
 \frac{i}{h} y_2
\end{array}\right)
  \left|
\left(\begin{array}{cc} i (A+\kappa^2/h) & i\kappa/h \\
i\kappa/h & i/h \end{array}\right)
\right)\right.
\end{align}
which results in a `transformed' $\tau$-matrix,
\begin{align}
\label{tautr}
\tau_{tr}=\frac{i}{\pi} \left(\begin{array}{cc}\log{\left( 
\frac{4\Phi^2}{\delta\Delta} \right)} &\log{\left(
\frac{4\gamma}{\delta}\right)}\\
\log{\left(\frac{4\gamma}{\delta}\right)} & 2 \log{\left( 
\frac{4\gamma}{\delta} \right)}\end{array}\right)  +
\frac{2\gamma}{\pi\Phi}\left(\begin{array}{cc}1 &1\\
1 & 0\end{array}\right).
\end{align}
Once the $\theta$-function has been written in terms of a 
$\tau$ matrix
with only large elements, its degeneration into trigonometric
functions is easily obtained from the definition of the
$\theta$-function in terms of an infinite sum,
\begin{align} \label{thetafunctionexpansion}
\theta \left( \vec{u} | \tau  \right) \equiv \sum_{\vec{m} \in
\mathbb{Z}^g} e^{ i\pi ( \vec{m} \cdot \tau  \vec{m} + 2
\vec{m} \cdot \vec{u}) }.
\end{align}
The dominant terms in the sum will come from the $\vec m$ 
close to the the stationary point of the real part of the 
exponent:
\begin{align}
\label{m0}
\vec{m}_0  =- (\mbox{Im}\tau)^{-1} \mbox{Im}\vec{u}
\end{align}
($\vec{m}_0$ is not necessarily a vector of integers). When 
$\vec{m}=(m,n)$
deviates much from $\vec{m}_0$ the exponential becomes 
rapidly
smaller because of the largeness of $\tau$, such that the
 sum is
dominated by only a few exponential terms -- a (hyper-) 
trigonometric
polynomial. Because $\tau$ is logarithmic in $\delta$ and 
appears
linearly in the exponentials, the sub-dominant terms in the 
sum will
be suppressed by powers of $\delta$.

\subsection{Recovering the Cooper Instability}

By setting $\Phi = \Delta_0/\sqrt{2}$ we completely suppress the 
gap, returning the system to the normal-metal state. In this limit 
we should see the absolute value of the order parameter execute 
the train of $\cosh^{-1}$ solitons found in previous work on the 
Cooper instability \cite{bls}. This behaviour corresponds to a 
simpler two-spin solution. This degeneration into a system 
described by fewer spins (or lower-genus $\theta$-functions) 
is expected whenever we close a branch cut on the Riemann 
surface for $Q(u)$ (here that joining the roots at $\pm i \Delta$),
 so long as the initial conditions are such that there is a $u_i$ 
 pinned at the resulting double root (here $u_1=0$) 
 \cite{kuznetsov}. This is the case when the perturbation is such 
 that $\Delta\ll\delta$, as is appropriate if the perturbation is most 
 significant near the Fermi surface\footnote{To see this, consider 
 the initial condition as deriving from a one-spin solution in which 
 $\delta=\Delta=0$ (here all the $s^{(i)}_-=0$, so that $B=0$) by 
 a perturbation which opens the branch cuts of size $\delta$ and 
 $\Delta$. After the perturbation, the (three-spin) $B(u)$ is $\sim 
 \delta (u-w_0)(u-w_1)$ for some $w_{0,1}$. Deducing the size 
 of the branch cuts in terms of $w_0$ and $w_1$ from the 
 expression for $Q(u)$ in terms of $A(u)$ and $B(u)$, and 
 demanding that $\Delta\ll\delta$, we see that one of the $w$ 
 is close to the $\Delta$-sized cut and one is close to the 
 $\delta$-sized cuts; as $\Delta\rightarrow 0$, the former root 
 is pinned to the origin.}. Note that here $\Delta\neq\Delta(0)$, 
 since the former does not include the effect of the perturbation. 
 With these initial conditions, an appropriate vector of constants 
 $\vec c$ is given by
\begin{align}
\label{ccooper}
\vec c =  \left(\begin{array}{cc}
 \frac{i}{2\gamma\Phi} \log{\left( \frac{4\gamma}{\delta}\right)}
 -\frac{1}{2\Phi^2}\left(\log  \left( \frac{4 \Phi^4}{\gamma\delta
 \Delta^2}\right)-2\right) ,
  & \frac{1}{\Phi} \end{array} \right).
\end{align}
Once we have expressions for the theta functions in terms of 
the `large' $\tau$-matrix, we extract the asymptotes in the 
manner described above. Consider one $\theta$-function, 
say that in the denominator of  (\ref{thetafunctionexpression}).
 To begin with we find the stationary point 
 $\vec m_0=(m_0,n_0)$ via (\ref{m0}), and see that while the 
 value of $n_0$ changes with time, $m_0 \simeq -1$ for small 
 $\Delta$. If we take only this $m$, we have a genus-one 
 $\theta$-function as expected on general grounds. The 
 denominator degenerates similarly, and the ratio has a 
 quasiperiod
 \begin{align}
 \label{period}
 t_0=\gamma^{-1} \log\left(\frac{4\gamma}{ \delta}\right)
\end{align}
corresponding to the quasiperiod of the genus two $\theta$-
functions. Numerator and denominator each have a 
stationary value of $n$, which can differ from that given by 
(\ref{m0}),
\begin{align}
n_0 (m)=-(\mbox{Im}(\tau_{22}))^{-1}\mbox{Im}(u_2+m
\tau_{12} )
\end{align}
and the argument of each sum is a Gaussian in $n$ whose 
width is fixed by the $(2,2)$ component of the transformed 
$\tau$ matrix.

For generic $t$ we can ignore all but one $n$ for the 
denominator, giving a single exponential, but when the stationary 
value of $n$ is close to halfway between two integers two values 
of $n$ are of comparable importance, yielding an expression for 
$\theta$ in terms of a trigonometric function. (The numerator 
behaves similarly half a period later, but at these points the ratio 
is insignificantly small.) The end result is a train of solitons 
separated by $t_0$, each of the form (to leading order):
\begin{align}
\label{coopergap}
\Delta(t)=\frac
{ 2 \gamma }
 {\cosh{(2\gamma t)}}.
\end{align}
The vector $\vec c$ and the overall normalization are fixed in 
the following way. The values of the $u_i$ (including $u_1=0$) 
determine $B(u)$ up to $\Delta(0)$: $B(u)=\frac{2\Delta(0)}
{\lambda}u (u-u_2) $. Equation (\ref{shiftedQ}) gives $Q(u)$ in 
the $\Delta\rightarrow 0$ limit. Writing $Q(u)$ in terms of $A(u)$ 
and $B(u)$, we see that $A(u)=\frac{2}{\lambda} u (u-w_1)
(u-w_2)$ for some $w_i$ that are either both real, or conjugate 
to each other. Matching the coefficients of $Q(u)$ written in terms
 of $A(u)$ and $B(u)$ with (\ref{shiftedQ}) gives a family of 
 acceptable solutions for $u_2 (0)$ and $\Delta(0)$, 
 corresponding 
 to different stages in the time evolution. Choosing a particular 
 $u_2(0)$ allows us to integrate to get the $\vec c$ above, and 
 also fixes $\Delta(0)$ and thus the normalization of our solution. 
 In the above formula we have omitted an overall phase rotation
  which corresponds to a redefinition of the chemical potential.

The form of the above soliton conforms exactly with previous 
results
 \cite{bls,bl2} for the oscillations of the order parameter following
  a
  sudden turn-on of the BCS interaction. Interestingly, it also 
  conforms
   exactly with the result of the next section, where we assume 
   $\delta\ll\Delta$, in the limit that $\Delta\ll\gamma$. This is 
   despite
    the difference in the relative sizes of $\Delta$ and $\delta$ 
    in the 
    two cases, which implies very different initial conditions for 
    the $u_i$.

\subsection{Main Case}

We now consider the case $\delta\ll\gamma,\Delta$.

\subsubsection{Initial Conditions for Small Perturbations}
An important difference between this case and the 
previous is the value of $\vec c$, which (\ref{c}) gives 
in terms of the initial positions of our variables $u_i(t=0)$. 
These are the zeroes of $B(u)$ at time $t=0$, which in 
the unperturbed case coincide with the zeroes of $G(u)$ 
as one can show using the self-consistency equation. For 
a small perturbation, the $u_i$ remain close to the branch 
cuts at $\Phi \pm i \gamma$; let us call them $u_\pm$. It 
is useful to define $z_\pm$ as the distance of $u_\pm$ 
from the centers of the branch cuts in units of $\delta$, 
which may be complex:
\begin{align}
u_{+}(t=0)=\Phi+ i\gamma+z_{+}\delta;
\qquad u_{-}(t=0)=\Phi- i\gamma+z_{-}\delta^*.
\end{align}
The integrals (\ref{c}) defining $\vec c$ can then be given 
through $z_\pm$ using the function $I_\pm$ defined as
\begin{align}
\label{Is}
I_\pm= - \int_{i}^{z_\pm}  \frac{\mathrm{d}w}{\sqrt{w^2+1}}=
 \frac{i\pi}{2}-\mbox{arcsinh} (z_\pm).
\end{align}
The expression for $c_i$ is then given by\footnote{The minus
 sign in front of the integral in (\ref{Is}) is due to the fact that
  the $u_i$ have to be chosen on the lower Riemann sheet 
  in this case. This can be ascertained by examination of the 
  final result once derived.}:
\begin{align}
c_1=\frac{
I_+ \left(1-i\frac{\gamma}{\Phi} \right)-I_- \left(1+i\frac{\gamma}
{\Phi} \right)
}{2i\gamma\Phi}, \qquad
c_2= \frac{ I_+-I_-}{2i\gamma};
\end{align}
up to corrections suppressed by $\gamma^2/\Phi^2$ and 
$z\delta/\gamma$.

The form of the solution is sensitive to the values of $c_1$ 
and $c_2$. But as noted above, these are not independent 
parameters. The necessary constraints can be found in the 
following way. We first find expressions for $z_{\pm}$ in 
terms of the perturbations to $A(u)$ and $B(u)$. Let $u_I$ 
be the position of a root of $B(u)$ before the perturbation 
is added, and $u_F$ its position afterward. Expanding 
$B=B_0+\delta B$ about the initial position of the root 
$u_I$ tells us that, due to the perturbation, $u$ travels a 
distance given by
\begin{align}
(u_F-u_I)\simeq -{\delta B(u_I)}/{B_0 '(u_I)}.
\end{align}
Similarly we can expand $Q\propto A(u)^2+B(u)B^*(u)$, 
taking into account the fact that it has a double root to 
begin with, and the fact that before the perturbation $A_0$
 and $B_0$ are related by $B_0(u)=\frac{\Delta}{\Phi+u}A_0(u)$. 
 This yields both the width of the branch cut (i.e. $2\delta$ or 
 $2\delta^*$) that is opened up by the perturbation, and the 
 location
  of its centre, in terms of $\delta B(u_I)$ and $\delta A(u_I)$. 
  We omit these formulas. Then the $z_\pm$ are given by
\begin{align}
z=\frac{(u_F-u_I)-(\textrm{displacement of centre of branch cut})}
{(\textrm{complex half-length of branch cut})}.
\end{align}
Defining $\delta B(u)=\frac{\Delta}{u} \delta A(u)+\delta C(u)$, all 
the $\delta A$s and $\delta B$s disappear in favor of $\delta C$s, 
and we can expand without worrying about the relative size of 
$\delta A$ versus $\delta B$:
\begin{align}
\label{z}
z_\pm= \frac{\Phi\pm i\gamma}{\Delta}
\sqrt{\frac{\delta C(\Phi\pm i\gamma)}{\delta C^*(\Phi\pm 
i\gamma)}}.
\end{align}
Since $\delta C^*(\Phi+i\gamma)=\delta C(\Phi-i\gamma)^*$, 
this tells us that $|z_+ z_-|=\Phi^2/\Delta^2$ to leading order, 
and that\footnote{For this last we must resolve a sign ambiguity 
coming from the square roots, which is most easily done by 
examining the special case where $\delta$ is real and the centers 
of the branch cuts do not move.} $\arg z_+/z_-$ is of order 
$\gamma/\Phi$. Eq. (\ref{z}) yields the following constraints on 
$c_i$ or $z_\pm$:
\begin{align}
p_1&\equiv  \mbox{Re }(c_1 \Phi^2 - c_2 \Phi)= \frac{1}{2}
\log 4 |z_+ z_-| =
\log \frac{2\Phi}{\Delta}, \\
p_2&\equiv\mbox{Re } (\gamma  c_2)=\frac{1}{2}\arg\frac{z_-}
{z_+}\simeq0.
\end{align}
These combinations of $c_1$ and $c_2$ are precisely those 
necessary for the correct expansion of the $\theta$-functions -- 
for example $p_1$ dictates which integers $(m,n)$ give the 
leading order contributions to the representation of the theta 
function as a sum (\ref{thetafunctionexpansion}).

Having derived these results by expanding $B(u)$, $A(u)$ and
 $Q(u)$, we must ask when they are valid. Assuming that 
 $ \sqrt{ \delta C(u_I) / \delta C^*(u_I)}$ is approximately of order
  one, we find that the roots of $B(u)$ move a distance of order 
  $\delta\Phi /\Delta$. This quantity must be much smaller than 
  $\gamma$, the scale on which our initial polynomials vary. So 
  a necessary condition for the validity of these approximations is
\begin{align}
\delta \ll \gamma \Delta / \Phi.
\end{align}
This excludes of course the Cooper instability case, where one
 root of $B(u)$ is a distance of order $\Phi$ from the start-points
  of the integrals in (\ref{Is}). Since $B(u)$ vanishes as 
  $\Delta\rightarrow0$, in this case it is not legitimate to assume 
  that $\delta B\ll B_0$.

\subsubsection{Form of the Solution}

Again we use the `transformed' $\theta$-functions, and extract 
the dominant exponentials from the sums defining them, 
(\ref{thetafunctionexpansion}). The precise values of $\vec c$ 
depend on the nature of the perturbation, but the information 
obtained above is enough to establish the nature of the 
solution up to (a) an overall shift in time and (b) a shift of the 
fast oscillations within their envelope. Up to such a shift, each 
soliton has the form (we neglect $\gamma/\Phi$ corrections):
\begin{align}
\label{gap2}
\Delta(t)=\Delta\left(
1+
\frac{2\gamma}{\Delta}e^{-2 i \Phi t}\text{sech} {(2\gamma t)}
\right)
\end{align}
and solitons occur at intervals of $t_0=\gamma^{-1}\log
(4\gamma/\delta)$.

More explicitly,
taking into account the expressions for the shifts in terms of 
$\vec c$, the first soliton has the form
\begin{align}
\label{gap3}
\Delta(t)= \Delta
\left(
1-
\frac{2i\gamma}{\Delta}
\exp{\left(-2 i \Phi t - c_2 \Phi + c_1 \Phi^2+\log\frac{\Delta}
{2\Phi} + i \arg \delta \right)}
\text{sech} {\bigg(2\gamma (t-t_0/2)-i\gamma c_2\bigg)}
\right).
\end{align}
From (\ref{Is},\ref{z}), $\mbox{Im }c_2$ is of order $\gamma^{-1}
\log(\Phi/\Delta)$, so that if $\delta$ is sufficiently small the first 
soliton takes about half a period to appear.

\section{Conclusion}
\label{conclusion}
We have found the short-time behavior of a BCS 
superconductor 
following a small
perturbation to the imbalanced initial conditions given by 
Eq.
\ref{npwithimbalance}. These initial conditions show a 
suppression 
of the gap \cite{galaiko} with increasing imbalance $\Phi$, 
and an 
instability \cite{spivak} when $\Phi>\Phi_c$ which becomes 
the 
celebrated Cooper instability when the gap is
fully suppressed. As $\Phi$ is increased beyond $\Phi_c$, the 
gap 
oscillations following upon the instability grow in magnitude. 
They 
take the form of a train of solitons, each of duration 
$\sim\gamma^{-1}$ and magnitude $\sim\gamma$, and 
containing 
oscillations on the shorter timescale $\Phi^{-1}$ (Eq. \ref{gap2}, 
and 
pictured in Fig. \ref{overView}). These oscillations
should be observable if appropriate initial conditions can be 
prepared 
in a controlled fashion. 

A stronger motivation for the work is that the oscillatory behavior 
is relevant to evolution on longer time scales 
($\sim\tau_\epsilon$) in experimental situations with large 
imbalance. In particular, the oscillations are relevant to the 
relaxation of the imbalance, which in the absence of the 
instability occurs at a rate which vanishes with the gap as 
$T\rightarrow T_c$ \cite{tinkham, schmidschon}. Understanding 
the short-time dynamics of (\ref{VolkovKogan}) is a first step; to 
determine quantitatively what happens on long time scales it is 
necessary to compute how
collisions modulate them. The moduli of the
solution, i.e. the variables used to parameterize the
kinds of short time behavior, can be taken to be the
roots of $Q(u)$. These roots or moduli vary
slowly with time on account of collisions. The non-equilibrium 
state
of the system at long times may correspond for example either 
to an unchanging
set of moduli, or to a limit cycle in moduli space. Such an analysis 
is beyond the
scope of the current paper, but will involve (\ref{gap2}) and 
possibly generalizations.

Our explicit expressions for the behavior of the order parameter 
apply when  $\delta\ll \gamma$. In this limit, where the solitons 
are widely spaced, the expressions simplify greatly; but the 
generalization to larger $\delta$ may be necessary to treat 
imbalance relaxation in realistic situations. As an idealized 
Gendankenexperiment, our limit can be realized by 
instantaneously injecting electrons at the Fermi level to a system
 at the instability point -- it can be shown using the definition 
 (\ref{specpol}) that such a perturbation increases the instability
  rate $\gamma$ while hardly increasing $\delta$\footnote{
  What is important is the structure of the Riemann surface 
  given by the spectral polynomial (\ref{specpol}), rather than 
  the details of the distribution of the spins.}. (A similar analysis
   shows that in a system with tunable interaction strength, 
   increasing the coupling of a system with $\Phi\lesssim\Phi_c$
    does the same.) The subsequent evolution of such systems 
    on time scales $\tau_\epsilon$ may increase $\delta$ further.

The present work neglects spatial inhomogeneities. Whether 
they change the picture qualitatively in systems larger than 
the coherence length remains to be investigated. Gap 
oscillations can also parametrically excite inhomogeneous 
modes, as shown in \cite{cooperturbulence} for the Cooper 
instability.

We have already mentioned that thermal processes act by 
slowly perturbing the dynamics considered. We have not 
mentioned thermal fluctuations in the initial conditions, which 
were analysed for the Cooper instability in 
\cite{bl2,developingpairing}. These fluctuations disappear when
 the level spacing, or the effective level spacing in a coherence 
 length, goes to zero, but will cause variations in the parameters 
 of the solution (e.g. our $\delta$) when it is finite. While the 
 evolution is of the same form in each realization, it was shown 
 that variations in the parameters between different realizations 
 are qualitatively important for averages (e.g. of the absolute 
 value of the gap) over realizations. Such averages would be 
 relevant to experiments involving direct observation of gap 
 fluctuations.
\\ \\
The authors would like to thank I.A. Gruzberg, L.S. Levitov 
and P.B. Wiegmann for discussions. In particular we are 
indebted to B. Spivak for indispensable advice and guidance.
\\
\\
EB was supported by grant number 206/07 from the ISF.

\end{document}